# Ultrafast photo-thermoelectric currents in graphene junctions in the mid-infrared

Nina Pettinger[1,2], Michel Panhans[3], Johannes Schmuck[1,2], Sebastian Loy[1,2], Xiaoyi Zhou[1,2], Chengye Dong,[4,5] J.A. Robinson[4,5,6,7,8], Sergey Zherebtsov[1,2], Christoph Kastl[1,2], Frank Ortmann[3*], Alexander W. Holleitner[1,2*]

[1] Walter Schottky Institute and Physics Department, Technical University of Munich, Am Coulombwall 4a, 85748 Garching, Germany

[2] Munich Center for Quantum Science and Technology (MCQST), Schellingstr. 4, 80799 Munich, Germany

[3] TUM School of Natural Sciences, Department of Chemistry and Atomistic Modeling Center, Munich Data Science Institute, Technical University of Munich, 85748 Garching, Germany.

[4] Two-Dimensional Crystal Consortium, The Pennsylvania State University, University Park, PA, USA

[5] Department of Materials Science and Engineering, The Pennsylvania State University, University Park, PA, USA

[6] Department of Chemistry, The Pennsylvania State University, University Park, PA, USA

[7] Department of Physics, The Pennsylvania State University, University Park, PA, USA

[8] Department of Engineering Science and Mechanics, The Pennsylvania State University, University Park, PA, USA

[*] Contact Email: holleitner@wsi.tum.de, Frank.Ortmann@tum.de

## Abstract

**Graphene is widely recognized for its ultrafast and broadband photocurrent response, but whether the broadband ultrafast characteristics are preserved at mid-infrared wavelengths with photon energies below the optical phonon energy remains an open question. Here, we investigate the carrier dynamics in graphene junctions under mid-infrared excitation using an ultrafast pump–probe photocurrent spectroscopy. We utilize dual split gate devices to demonstrate that the photo-thermoelectric effect can dominate the photoresponse of graphene also for a mid-infrared femtosecond excitation. We observe that graphene retains its broadband photocurrent response in this spectral region, but the photocurrent relaxation time increases from ~2 ps below 8-9 μm up to 3 ps at longer mid-infrared wavelengths. The absence of a pronounced phonon bottleneck in the decay dynamics at room temperature suggests an efficient interplay of electron-electron and electron-phonon scattering even for photon energies below the optical phonon energy in graphene. The observed wavelength dependence of the photocurrent relaxation times is consistent with energy-dependent theoretical relaxation times as derived from a microscopic transport theory of graphene that includes electron-phonon coupling within a Holstein-Peierls Hamiltonian.**

## Introduction

Graphene is an attractive platform for broadband optoelectronics because it uniformly absorbs light across an exceptionally broad spectral range[1]. Therefore, unlike conventional semiconductors, graphene-based devices are not restricted to the visible or near-infrared: their operation naturally extends into the THz and mid-infrared (mid-IR), where applications in chemical sensing, free-space communications, and thermal imaging demand compact, ultrafast and sensitive detectors[2,3]. To date, however, most demonstrations of ultrafast graphene photodetectors have focused on the near-infrared spectral range, reporting picosecond-scale response times[4–13]. Extending such performance into the mid-IR spectral range remains largely unexplored.

At a fundamental level, understanding the relaxation processes of photoexcited carriers is essential for optimizing such graphene-based optoelectronic devices. The general picture of carrier excitation and relaxation[14–23] in graphene can be summarized in the following way. Upon light absorption, electrons in graphene are driven out of equilibrium into a non-thermal distribution. Next, the excited electrons and holes thermalize into an ensemble of hot carriers, usually via fast electron-electron and Auger interactions (less than 100 fs)[21–24]. For slightly or undoped graphene, additionally, interband impact ionization processes play a role and can lead to a significant carrier multiplication in the conduction band[21]. These hot carriers then dissipate their excess energy through interactions with the lattice, which is predominantly considered to be a defect-mediated coupling to acoustic phonon[25] (or super-collision) cooling[26] and a direct coupling to optical phonons[23]. While the optical phonon emission process itself can generally be considered very efficient, with electron-phonon scattering times on the order of or below 100 fs, the overall cooling rate appears to be intrinsically limited by a so-called hot phonon bottleneck, which slows down the equilibration of the hot optical phonons with the lattice via anharmonic decay into acoustic phonons, and thereby the overall electron cooling rate, to several ps[27]. Moreover, the large optical phonon energies in graphene (on the order of 160 - 200 meV) open the possibility of a second considerable cooling bottleneck in undoped graphene, in particular for a mid-IR photoexcitation (a photon energy of 160 meV in vacuum corresponds to a wavelength of 7.7 µm): when the excess energy of the excited, thermalized electrons becomes comparable to the optical phonon energy, their cooling dynamics are expected to slow down significantly[28], with reported relaxation times on the order of 300 ps at low temperatures[23]. While such a bottleneck is considered advantageous for a possible hot carrier extraction in solar cell devices, it would be an intrinsically limiting factor for ultrafast mid-IR to THz detectors. Lastly, in graphene encapsulated with hexagonal boron nitride (hBN), as is common for high-mobility and high-performance devices, additional cooling routes may emerge by interactions with the surrounding polar dielectric, such as coupling to surface polar phonons or hyperbolic phonon polaritons[29–31]. These highly confined, long-lived infrared modes may efficiently accept energy from hot carriers in the adjacent graphene layer, thereby offering remote, substrate-mediated cooling pathways that compete with the more intrinsic optical phonon emission[32,33].

Here, we investigate the ultrafast photocurrent generation in graphene junctions at room temperature and under a mid-IR excitation (5 µm – 12 µm). Using a combination of time-integrated and time-resolved photocurrent measurements, we find that the mid-IR response of the graphene devices is governed by photo-thermoelectric currents, which maintains an ultrafast broadband photoresponse with photocurrent relaxation times on the order of a few picoseconds.

## Results

### Dual-gated graphene devices

We fabricate graphene-based p-n junction devices on a silicon/silicon dioxide (Si/SiO$_2$) substrate and probe the carrier dynamics by a mid-IR, time-resolved photocurrent autocorrelation scheme (Fig. 1a, see methods). The junction is defined by a pair of platinum (Pt) split-gates. A hexagonal boron nitride (hBN) flake transferred onto the gates acts as a dielectric spacer. A monolayer graphene sheet with source-drain contacts serves as the active material. Figure 1b shows an optical image of the final device structure with the different layers highlighted in false color. For all photocurrent measurements, we apply a bias voltage $V_{sd}$ at the source contact, and we measure the (photo)current by a transimpedance amplifier at the drain contact as function of gate voltages $V_{g1}$, $V_{g2}$ on the two split-gates. The two-terminal device resistance without any laser excitation as function of the common gate voltage $V_g = V_{g1} = V_{g2}$ (Fig. 1c) and the individual gate voltages $V_{g1}$ and $V_{g2}$ (Fig. 1d) shows a clear resistance peak, which we assign to the Dirac point of the graphene. Notably, we find negligible hysteresis upon forward and reverse voltage sweeps demonstrating a high quality of the gate dielectric (Fig. 1c).

### Mid-infrared photocurrent autocorrelation

For the photocurrent measurements, we focus a chopped, mid-IR laser beam onto the center of the device (white cross in Fig. 1b). The resulting, time-integrated photocurrent is resolved with a standard lock-in detection (see methods and

Supplementary Fig. 1 for a photocurrent map). Figure 1e shows the gate-dependence of the photocurrent at $V_{sd}$ = 0 V and at an excitation wavelength of $\lambda_{photon}$ = 8 µm revealing a distinct six-fold pattern, which is a characteristic signature of the photo-thermoelectric effect in graphene *p-n* junctions[13,34–37].

To resolve the underlying carrier dynamics, we use a time-resolved, pump-probe photocurrent autocorrelation scheme[4–11]. Figure 1f shows the photocurrent normalized to the excitation power as a function of the delay time $\Delta t$ between the pump and probe pulses. The time-resolved photocurrent exhibits a dip around zero-time delay, which is consistent with the expected behavior of the photocurrent autocorrelation for the case of sub-linear power scaling of the individual photocurrent pump- and probe-induced currents (cf. Supplementary Fig. 2). Near zero-time delay, we observe photocurrent oscillations due to the constructive and destructive interference of the co-polarized pump and probe beams on the sample[38]. We model the decaying photocurrent dynamics by an exponential function using:

$$I_{photo} = A_0 \cdot \exp(-|\Delta t|/\tau) + I_{offset}, \quad (1)$$

with $A_0$ the amplitude at zero-time delay ($\Delta t$ = 0) and $\tau$ the characteristic photocurrent relaxation time. The term $I_{offset}$ accounts for the background current, which is present because we demodulate the photocurrent only at the chopping frequency of the pump beam (see methods). Additionally, we include a linear background in the fitting procedure that accounts for slow drifts during the measurement.

We note that we reproduced the ultrafast photo-thermoelectric currents on another graphene *p-n* junction device (Supplementary Fig. 3). Moreover, graphene devices on a single $SiO_2$ back gate show the ultrafast mid-IR response as well (Supplementary Fig. 4). Supplementary Table 1 summarizes all seven samples studied within the scope of this work.

**Mid-infrared wavelength dependence**

Figure 2 presents the wavelength- and gate voltage-dependent fitting parameters extracted from the ultrafast photocurrent measurements. For each configuration of split gate voltages ($V_{g1}$, $V_{g2}$), an ultrafast photocurrent measurement as a function of delay time (Fig. 1f) is analyzed. The fit results, using the fit function in Eq. 1, are shown for the wavelengths 5 µm – 12 µm. Specifically, we map the amplitude $A_0$ (Fig. 2a), the photocurrent relaxation time $\tau$ (Fig. 2b), and the offset current $I_{offset}$ (Fig. 2c). Note that if the photocurrent amplitude $A_0$ does not exhibit a clear dip or peak, we cannot extract a meaningful value for $\tau$. Such data points are set to be white in Figs. 2a and 2b. The offset current $I_{offset}$, which corresponds to the time-integrated photocurrent, reproduces the characteristic six-fold pattern of the photo-thermoelectric effect for the full range of the investigated excitation wavelengths (5 µm – 12 µm, Fig. 2c). The time-resolved amplitude $A_0$ (Fig. 2a) follows a similar pattern but sign-reversed: if $I_{offset}$ is positive, a dip in $I_{photo}(\Delta t)$ (cf. Fig. 1f) will correspond to $A_0$ being negative, and vice versa, which agrees with the expected behavior for a power-saturated, sub-linear photocurrent (cf. Supplementary Fig. 2). We find this anti-correlation holds well when the junction is biased into the *n-p* or *p-n*-regime (cf. Figure 1e for the different doping regimes). Notably, when the junction is biased into the *n-n*- or *p-p*-regime, we sometimes observe a peak-shaped rather than dip-shaped time-resolved photocurrent (cf. Supplementary Figs. 2 and 5)[5].

Our mid-IR pump-probe experiments allow us to connect the offset photocurrent $I_{offset}$, which would be the quantity measured in a time-averaged measurement, to the ultrafast photocurrent dynamics via the time-resolved amplitude $A_0$. Their close correspondence (cf. Fig. 2) suggests that the photocurrent relaxation in our devices under mid-IR excitation is governed by the relaxation dynamics of hot carriers, which occur within a few ps similar to the case of near-IR excitation.[4–13]

**Mid-infrared optical absorption**

Both $A_0$ and $I_{offset}$ decrease with increasing wavelength up to 10 µm, except for a pronounced peak at 11 µm. To better highlight this behavior, Fig. 3a depicts the mean amplitude $A_{0,mean}$ with the values extracted from the top-left quadrant of Fig. 2a, i.e. the *n-p* regime, for each wavelength. Only non-zero entries, those measurements where a well-defined dip or peak is observed, are included in the average. The shaded error bands correspond to the error of $A_{0,mean}$. From this analysis, we confirm that the increase of $A_0$ peak at 11 µm is indeed statistically significant. Figure 3b summarizes the corresponding photocurrent relaxation time $\tau_{mean}$ averaged from the data in Fig. 2b. The shaded areas denote again the estimate of the standard error for $\tau_{mean}$. Specifically, for wavelengths between 5 µm and 9 µm, the dominant optoelectronic relaxation time $\tau_{mean}$ remains nearly constant at ~2 ps (Fig. 3b and Supplementary Fig. 6), indicating a fast and wavelength-independent cooling pathway of photo-excited charge carriers in this spectral range. Beyond 9 µm, however, the photocurrent relaxation time increases steadily, suggesting a slowdown of the dominant cooling mechanism of the photo-excited charge carriers. We note that the wavelength dependence of the decay time $\tau_{mean}$ is consistent across the different measured device architectures: it is observed in both dual-gate devices (Fig. 3b and Supplementary Fig. 6h) as well as in a single-gate device (Supplementary Fig. 6g).

To understand the wavelength dependence of the photocurrent amplitude, we calculate the absorption properties of the full heterostructure stack (Fig. 3c) using the transfer matrix method (Supplementary Note 1)[39,40]. The calculated absorbance (Fig. 3d) decreases monotonically up to approximately 9 µm, but then it shows a pronounced peak near 10 µm, in agreement with our experimental data (Fig. 3a). The increased absorption at this wavelength can be understood by resonances in the overall heterostack.

**Microscopic transport theory**

To rationalize the measured photon-energy dependence of the decay times, we turn to a microscopic transport description of charge carriers in graphene that explicitly accounts for electron–phonon coupling and that allows us to extract energy-dependent relaxation times within linear-response theory. We describe the charge transport over a range of energies accessed by mid-IR excitation by explicitly propagating electron-phonon wave packets in the presence of electron–phonon coupling to optical phonon modes. This provides access to carrier relaxation processes and possible phonon-dressing effects upon excitation into the conduction band (Fig. 4a).

Interband transitions across the Dirac point create a non-equilibrium distribution of charge carriers with electrons at energies $E = 1/2 \cdot \hbar\omega$ approximately above the CNP (Fig. 4a). To determine the characteristic relaxation time at these energies, we evaluate the energy-dependent transport scattering time from the dc conductivity in the zero-temperature limit. The underlying microscopic description of graphene is based on a Holstein–Peierls Hamiltonian that explicitly includes the coupling of charge carriers to optical phonon modes. The corresponding relaxation time as function of energy [41] is obtained as:

$$\tau(E) = 2 \frac{\lim_{T \to 0} \sigma_{\alpha\alpha}^{\mathrm{dc}}(E,T)}{e^2 D(E) v_{\mathrm{F}}^2}, \qquad (2)$$

where $\sigma_{\alpha\alpha}^{\mathrm{dc}}$ is the diagonal component of the dc-conductivity tensor, $D(E)$ the density of states of the Hamiltonian, and $v_{\mathrm{F}}$ the bare Fermi velocity of graphene at the Dirac point. Within this approach, the dc-conductivity is obtained from the time dependence of the electronic mean-square displacement $\Delta x_a^2(t)$, which is obtained by explicitly propagating electron–phonon wave packets in real time. The propagation is performed using a highly efficient numerical charge-transport method that fully accounts for electron-phonon interaction with optical phonon modes[42–44] (see Methods and Supplementary Note 3 for details).

As shown in Fig. 4b, this approach yields a pronounced energy dependence of the carrier relaxation time [Eq. (2)]. For reference, the dashed gray line indicates the parametric inelastic time scale $\tau_{\mathrm{ref}}$ at which the transport dynamics are investigated. In the absence of electron-phonon coupling, electrons propagate ballistically as reflected by a quadratic time dependence of the mean-square displacement $\Delta x_a^2(t) \propto t^2$ approximately equal for all energies *E* (black line). When the electron-phonon coupling is included, the relaxation time exhibits a marked reduction of the conductivity and hence *τ(E)* at energies around *E* ≈ 0.1 eV, followed by a gradual recovery toward the ballistic limit at higher energies of ~0.3 eV (blue line). Within the present model, this reduction originates from a coherent dressing of the electronic states by optical phonons that leads to a renormalization of the band velocity of about ~70%. For the parameters used in our simulation, this polaronic dressing occurs on ultrafast scales (~20 fs), which corresponds to the timescale of the optical mode energies. We note that this renormalization value is consistent with previous angle-resolved photo-emission spectroscopy (ARPES) studies on velocity renormalization in graphene[45] and it is significantly larger than predictions using electron self-energy renormalization schemes leading to a velocity renormalization of 4% to 8%.[47] The (dimensionless) EPC constant used in this work is 0.159 (see Supplementary Note 2). Previous TR-ARPES studies report an about 15% larger value of 0.187.[46] Note that the experimental reference reports the squared EPC constant (0.035). Therefore, the present microscopic theory captures independent experimental observations in several key aspects and therefore provides a reliable basis for explaining the mid-IR- photocurrent measurements.

# Discussion

We first discuss the extent to which the microscopic transport calculations capture the experimentally observed photocurrent dynamics. As shown in Fig. 4b, the calculated energy dependence of the theoretical carrier relaxation time covers the overall trend of the experimental photocurrent data very well (red crosses), including the pronounced minimum at intermediate energies. Assuming $\tau_{\mathrm{ref}}$ to be 3.7 ps, the calculation yields relaxation times that are comparable in magnitude to the experimentally observed values of the photocurrent relaxation time and correctly capture their photon-energy dependence. This agreement suggests that, also under mid-infrared excitation, the interaction of hot charge carriers with optical phonons plays a central role in governing the photocurrent relaxation dynamics in graphene. Generally, the assumed value of $\tau_{ref}$ (3.7 ps) is on the same order as reported for purely optical mid-IR pump-probe experiments at similar photon energies at room temperature[23]. However, a one-to-one comparison of time scales as deduced from purely optical and photocurrent experiments is beyond the current work.

We further note that the presented ultrafast mid-IR photocurrent experiments with an electric detection are in between optical pump/probe experiments[20–24] and lateral electrical transport experiments without an optical illumination conducted at the Fermi energy, where interactions with defects, charge impurities, and inhomogeneities of the sample can be dominating[48–50]. Importantly, our samples show a considerable room temperature mobility of ~25000 $cm^2V^{-1}s^{-1}$, such that we can assume to be in a regime where the cooling is not due to defect mediated processes such as supercollisions, but rather due the intrinsic optical phonon interactions and dynamics[14].

We finally point out that for a photon energy below twice the value of $E_F$, it is expected that the carrier distribution can be heated via free-carrier absorption processes and experimentally, that the relaxation time further increases[23]. For this purpose, we investigated the dual-gated devices also further away from the charge neutrality point ($V_{g1}$ and $V_{g2}$ = ±10 V, cf. Supplementary Fig. 8), corresponding to a Fermi energy of 0.3 eV, which is above the investigated photon energies (Fig. 2). However, we observe that the ultrafast mid-IR response vanishes entirely. This behavior is consistent with our comparison sample of epitaxial graphene on SiC, which has a Fermi energy of 0.4 eV and likewise shows no ultrafast mid-IR signal within the given signal to noise ratio (cf. Supplementary Fig. 9). Future experiments might be able to resolve such photocurrent dynamics at even lower photon energies and/or a higher Fermi-energy.

## Methods

**Device fabrication.** The devices were fabricated on Si wafers with a 285 nm thermally grown $SiO_2$ layer. Split gates were patterned on the substrate using optical lithography and metal evaporation of 3 nm platinum (Pt) followed by lift-off in acetone. A ~60 nm thick flake of hexagonal boron nitride (hBN) was mechanically exfoliated from bulk crystals and transferred onto the split gates using a dry transfer technique. Monolayer graphene, obtained by mechanical exfoliation from natural graphite, was identified by optical contrast and Raman spectroscopy, and subsequently transferred on top of the hBN. Source and drain contacts were defined by maskless UV-lithography, followed by deposition of 5 nm Ti and 150 nm Au using electron-beam evaporation, and lift-off in acetone. For the comparison sample of epitaxial graphene (EG), the graphene was prepared on silicon carbide (SiC) via thermal decomposition of SiC. Clean SiC substrates were first etched in 10% $H_2$/Ar mixture at 1400 °C and 700 Torr for 30 min. The substrates were then heated to 1800 °C in pure Ar at 700 Torr for 20 min to form monolayer EG.

**Electrical measurements without photoexcitation.** Dark electronic transport measurements were performed at room temperature in a nitrogen-purged environment. The gate voltages were applied to the split gates via a source-measure unit (SMU), and the resulting current through the graphene channel was recorded with a multimeter in combination with a transimpedance amplifier. For two-dimensional transfer maps, both split gate voltages were swept simultaneously over the desired range while recording the current.

**Photocurrent measurements with only the pump-pulse.** Photocurrent measurements were performed by focusing a tunable mid-IR laser beam onto the center of the graphene using an off-axis parabolic mirror. The laser beam had a diameter of about 50 μm at the plane of the sample, and it was modulated at 537 Hz using an optical chopper. The laser is the same as the pump-beam for the pump-probe measurements. The resulting photocurrent was detected with a transimpedance amplifier coupled to a lock-in amplifier, referenced to the chopper frequency.

**Ultrafast pump–probe measurements.** For time-resolved photocurrent measurements, the mid-IR beam was split into a pump and a mechanically delayed probe arm using a beamsplitter. The beams were recombined and focused collinearly onto the graphene channel. The photocurrent as a function of pump–probe delay was recorded with the same lock-in detection scheme as described above.

### Theoretical Methods

We use a highly efficient linear-scaling charge-transport approach based on the Kubo formalism to calculate carrier-relaxation times from dc-electron conductivities using the electronic mean square displacement via

$$\sigma_{\alpha\alpha}^{dc}(T) = \frac{e^2}{V}\frac{\beta}{2}\lim_{t\to\infty}\frac{d}{dt}\Delta x_\alpha^2(t). \tag{M1}$$

The electron-phonon coupling to the optical modes is implemented using the Holstein-Peierls Hamiltonian

$$\hat{H} = \sum_{\langle M,N\rangle}\gamma \hat{c}_M^\dagger \hat{c}_N + \sum_{\langle M,N\rangle \mathbf{Q}}\hbar\omega_\mathbf{Q} g_{MN}^{\mathbf{Q}}(\hat{b}_\mathbf{Q}^\dagger + \hat{b}_{-\mathbf{Q}})\hat{c}_M^\dagger \hat{c}_N + \sum_\mathbf{Q}\hbar\omega_\mathbf{Q}\left(\hat{b}_\mathbf{Q}^\dagger \hat{b}_\mathbf{Q} + \frac{1}{2}\right), \tag{M2}$$

with the nearest-neighbor electronic-coupling parameter $\gamma$, the phonon energies $\hbar\omega_Q$ with the combined index $\mathbf{Q} = (\mathbf{q}, \lambda)$, where $\mathbf{q}$ is the phonon wave vector and $\lambda$ is the mode index. The electron-phonon coupling (EPC) is described by the coupling constants $g_{MN}^{\mathbf{Q}}$. In the calculations of the relaxation time, a simplified version of the Holstein-Peierls Hamiltonain in Eq. (M2) is realized in the combined electron-phonon product Hilbertspace on a real-space grid of the graphene lattice and the corresponding mode patterns of the optical modes at the $\Gamma$-point of the Brillouin zone. A full description of the theoretical model and the calculated transport properties is included in the Supplementary Note 2.

## Data availability

The data as discussed in the current study is available from the authors upon reasonable request.

## Acknowledgements

We gratefully acknowledge the German Science Foundation (DFG) for financial support via the clusters of excellence MCQST (EXS-2111) and e-conversion (EXC 2089, project number 390776260), and the priority program 2244 (2DMP) via HO3324/13-2 and KA 5418/1-2 as well as the Munich Quantum Valley (K1). C.K. aknowledges funding through the European Union's Horizon Europe research and innovation programme under grant No 101076915 (2DTopS). C.D and J.A.R are supported by 2DCC-MIP under NSF cooperative agreement DMR-2039351 and the Penn State MRSEC Center for Nanoscale Science via NSF award DMR2011839.

## Author contributions

N.P., J.S., S.L., X.Z. performed the experiments, N.P., C.D., J.A.R., S.Z., C.K., A.W.H. conceived the experiments, M.P., F.O. performed the microscopic modelling and simulations. All authors wrote and read the manuscript.

## Competing interests

The authors have no competing interests.

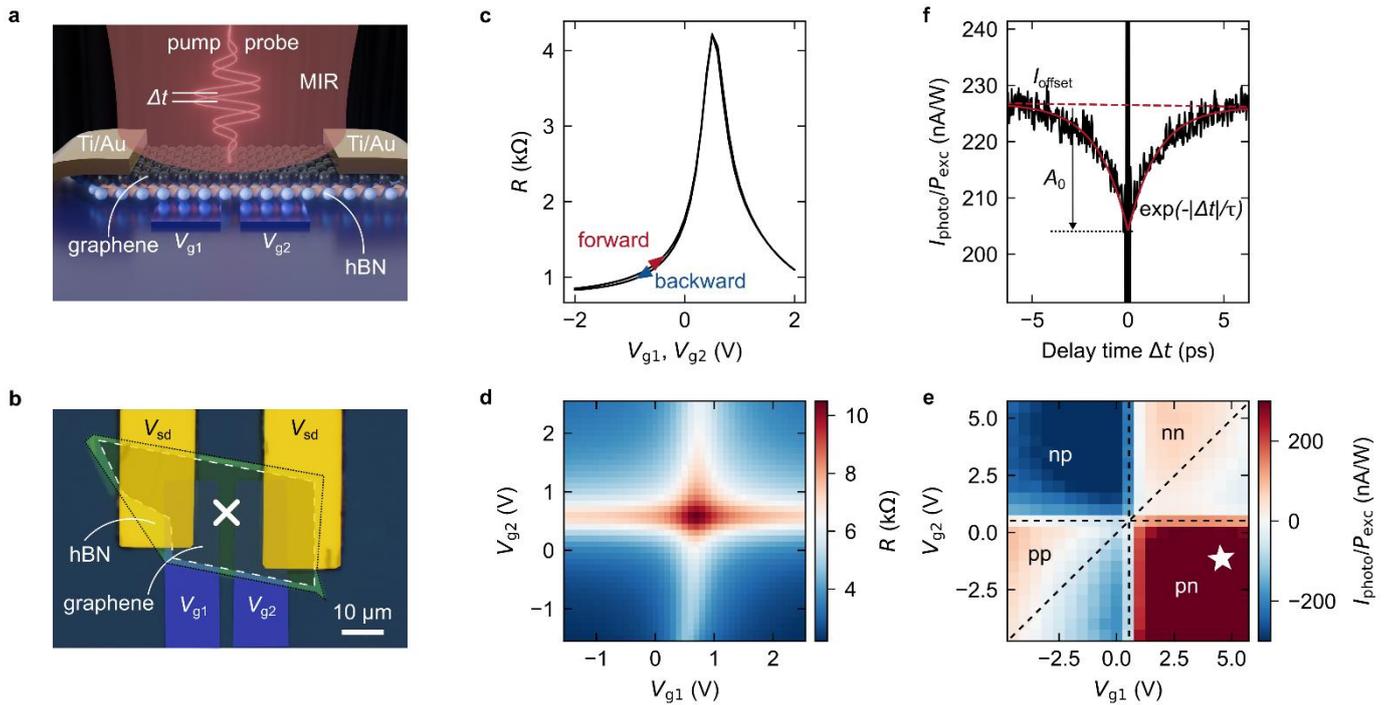

**Figure 1. Photo-thermoelectric response of dual-gated graphene.** (a) Experimental scheme with a graphene-hBN heterostack contacted by two gold electrodes on top of two platinum split gates. The latter allow applying the gate voltages $V_{g1}$ and $V_{g2}$ to the heterostack. The graphene is excited by mid-infrared (mid-IR) pump and probe pulses with a time-delay $\Delta t$ in-between. Bias voltage $V_{sd}$ is zero for all shown pump-probe experiments. (b) Optical image of a graphene-hBN heterostack with split gates (highlighted in blue), ~60 nm thin hBN (highlighted in green), monolayer graphene (highlighted in gray with white dashed outlines), and gold contacts (highlighted in yellow). White cross indicates the laser position. (c) Two-terminal resistance $R$ of graphene when varying both gates simultaneously. (d) $R$ vs. $V_{g1}$ and $V_{g2}$. (e) Time-integrated photocurrent $I_{photo}$ vs. $V_{g1}$ and $V_{g2}$ for using just the pump laser at an excitation wavelength of $\lambda_{photon} = 8$ μm. The laser is positioned at the center of the device [cf. white cross in Fig. 1(b)]. (f) Ultrafast photoresponse $I_{photo}$ vs. $\Delta t$ at $\lambda = 8$ μm, normalized to the excitation power $P_{exc}$. In red: exponential fit to equation (1). The values of $V_{g1}$ and $V_{g2}$ are indicated with a white star in Fig. 1(e).

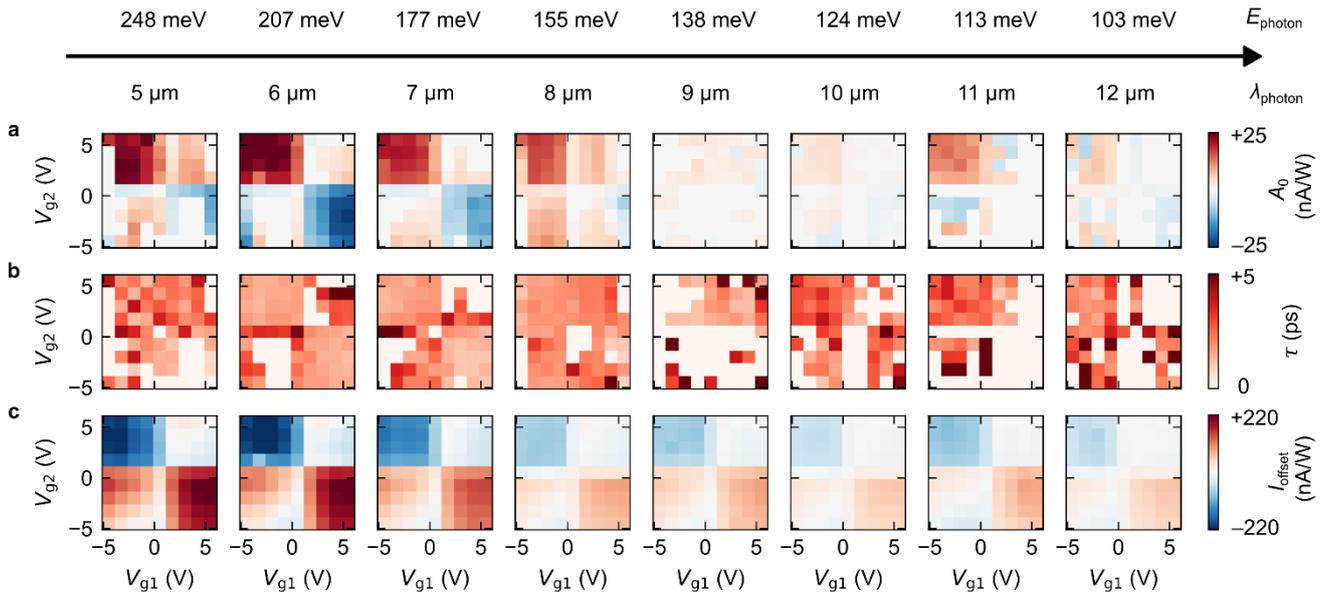

**Figure 2. Mid-IR wavelength dependence of ultrafast photo-thermoelectric response.** Ultrafast photocurrent contributions are measured for wavelengths 5 µm – 12 µm as in Fig. 1(f). (a) Fitted amplitude $A_0$ vs. $V_{g1}$ and $V_{g2}$. Values are normalized to the laser power per wavelength. (b) Fitted decay time $\tau$ vs. $V_{g1}$ and $V_{g2}$. (c) Fitted offset current $I_{offset}$ vs. $V_{g1}$ and $V_{g2}$. Top panel shows the utilized mid-IR wavelength (5 – 12 µm) and corresponding photon energies for each of the columns below.

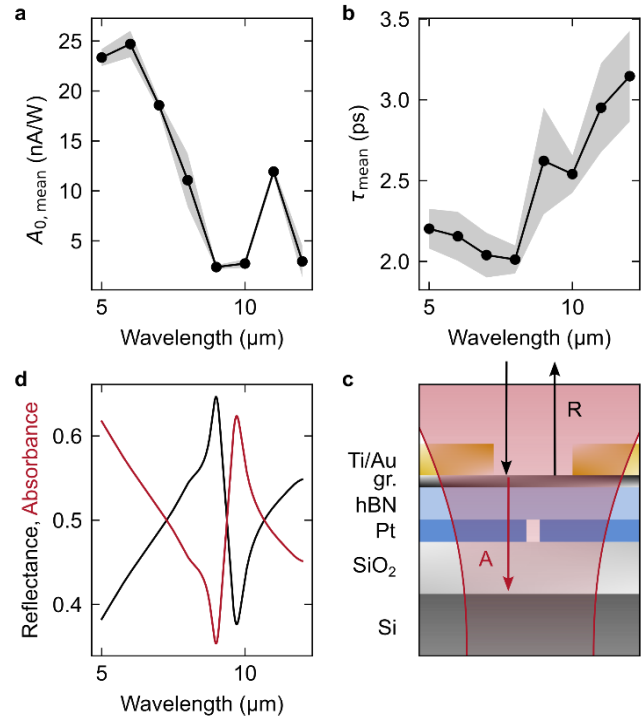

**Figure 3. Wavelength-dependent optoelectronic characteristics of the dual-gated graphene heterostack.** (a) Mean value of the fitted amplitude $A_0$ vs. photon wavelength. (b) Mean value of the fitted relaxation time $\tau$ vs. photon wavelength. (c) Schematic side-view of the overall heterostack including the Si/SiO$_2$-substrate, which allows to calculate the absorption $A$ and reflection $R$ of the incident light. (d) Calculated reflectance (black) and absorbance (red) of the overall heterostack as shown in Fig. 3(c) vs. photon wavelength. See text for details.

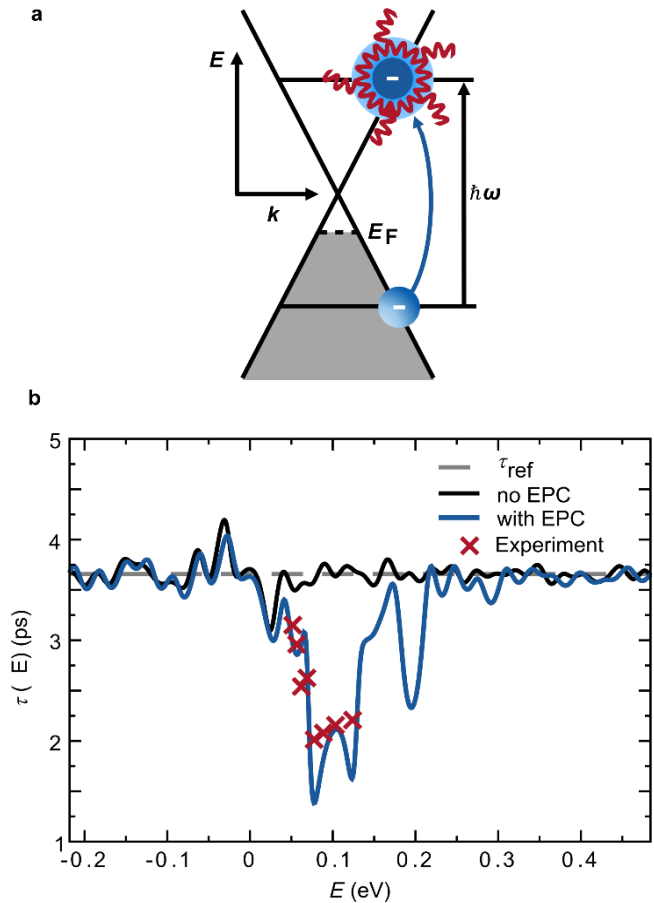

**Figure 4. Relaxation, hot carrier, and polaron effects in graphene.** (a) Schematic illustration of an excited electron that is coherently dressed with optical phonons leading to polaron dressing at time scales above several tens of fs. (b) Energy dependence of the experimental decay time (red crosses) compared to theoretical carrier relaxation times according to Eq. (2), including the case for no electron-phonon coupling (no EPC, black) and the interacting case (with EPC, blue), where the results of LO and TO optical phonons are averaged. The energy axis is half the photon energy used in the experiment as sketched in (a). The dashed gray line represents the total simulation time for inelastic processes serving as a reference time scale. Note that all curves are aligned with their Dirac point at $E = 0$.

# Supplementary Information

# Ultrafast photo-thermoelectric currents in graphene junctions in the mid-infrared

Nina Pettinger[1,2], Michel Panhans[3], Johannes Schmuck[1,2], Sebastian Loy[1,2], Xiaoyi Zhou[1,2], Chengye Dong,[4,5] J.A. Robinson[4,5,6,7,8], Sergey Zherebtsov[1,2], Christoph Kastl[1,2], Frank Ortmann[3*], Alexander W. Holleitner[1,2*]

[1] Walter Schottky Institute and Physics Department, Technical University of Munich, Am Coulombwall 4a, 85748 Garching, Germany

[2] Munich Center for Quantum Science and Technology (MCQST), Schellingstr. 4, 80799 Munich, Germany

[3] TUM School of Natural Sciences, Department of Chemistry and Atomistic Modeling Center, Munich Data Science Institute, Technical University of Munich, 85748 Garching, Germany.

[4] Two-Dimensional Crystal Consortium, The Pennsylvania State University, University Park, PA, USA

[5] Department of Materials Science and Engineering, The Pennsylvania State University, University Park, PA, USA

[6] Department of Chemistry, The Pennsylvania State University, University Park, PA, USA

[7] Department of Physics, The Pennsylvania State University, University Park, PA, USA

[8] Department of Engineering Science and Mechanics, The Pennsylvania State University, University Park, PA, USA

[*] Contact Email: holleitner@wsi.tum.de, Frank.Ortmann@tum.de

The PDF file includes:

Supplementary Note 1-3

Supplementary Figs. 1 to 10

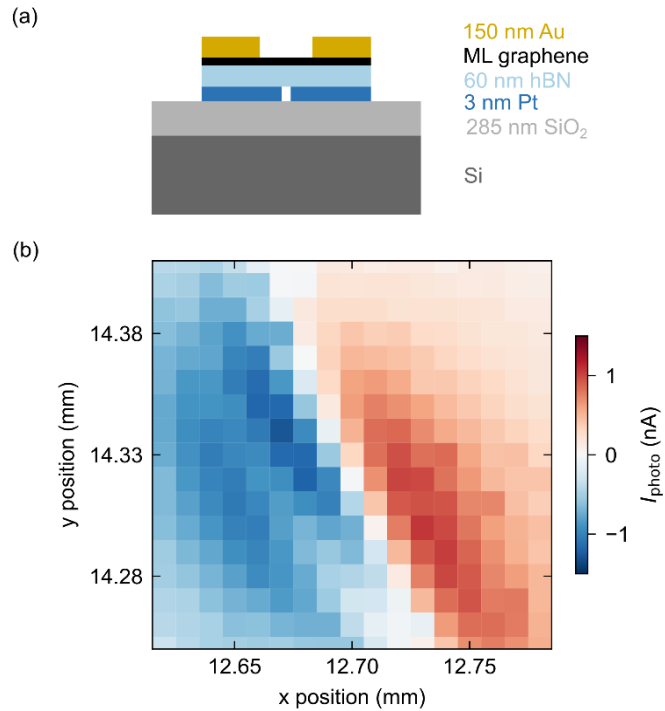

**Supplementary Figure 1. Split gate sample Graphene 1.** (a) Schematic side-view of the sample geometry. (b) Spatially resolved photocurrent (PC) map ($\lambda$ = 8 µm and $V_{g1} = V_{g2} = V_{sd}$ = 0 V). Each pixel length equals 10 µm.

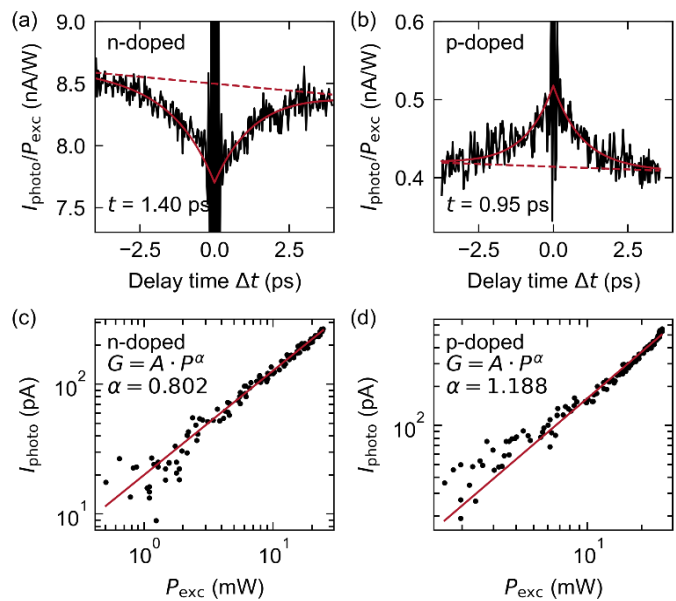

**Supplementary Figure 2. Power dependence of the photocurrent and corresponding photocurrent autocorrelation.** (a), (b) Exemplary ultrafast photocurrent measurements for (a) *n*-doping and (b) *p*-doping. (c), (d) Double-logarithmic plot of the photocurrent as a function of excitation power for (c) *n*-doping and (d) *p*-doping. We observe that for *n*-doping (*p*-doping) the signal shows a dip (peak) and a sublinear (superlinear) power dependence with an exponent of 0.802 (1.188) in accordance with literature[1,2].

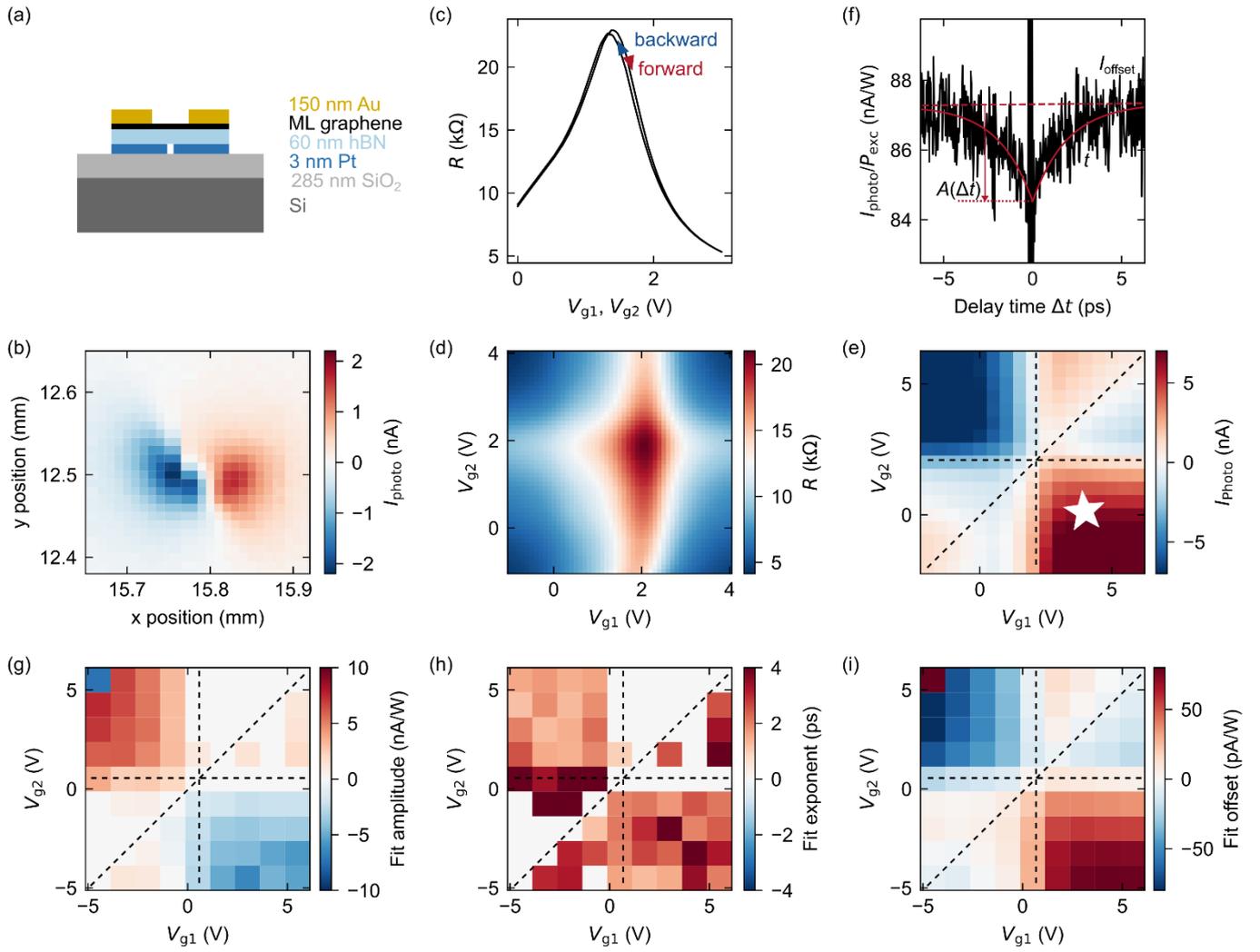

**Supplementary Figure 3. Sample 2 (split gate sample with graphene).** (a) Schematic side-view of the sample geometry. Identical to the sample geometry of sample Graphene 1 (main part). (b) Spatially resolved photocurrent (PC) map. Each pixel length equals 10 µm. Taken with $\lambda$ = 8 µm and $V_{g1} = V_{g2} = V_{sd}$ = 0 V. (c) Resistance $R$ of graphene when varying both split gate voltages simultaneously. (d) Resistance $R$ map of graphene when varying both split gates individually. (e) Photocurrent $I_{Pphoto}$ map when varying both split gate voltages individually, with $\lambda$ = 8 µm and $V_{sd}$ = 0 V (f) Ultrafast photo response $I_{photo}$ as a function of delay time $\Delta t$ between pump and probe pulse, normalized with the excitation power $P_{exc}$, in black. Exponential fit $A(0)\ exp(-|\Delta t|\ /\ t) + I_{offset}$ in red. (g)-(i) Fit parameter maps. At each split gates point ($V_{g1}$, $V_{g2}$), an ultrafast photocurrent measurement as a function of delay time is taken. The fit results are shown: (g) amplitude $A(\Delta t = 0)$, (h) exponent $t$, (i) offset $I_{offset}$.

| | Fabrication | Gate | Substrate | $n_{2D}$ | Ultrafast MIR response | $t$ (ps) | hBN thickness |
|---|---|---|---|---|---|---|---|
| Graphene 1 | mech. exfol. | split gates | Si/SiO$_2$ (285 nm) | dep. on $V_{g1}$, $V_{g2}$ | yes | 1.55 +/- 0.21 | 60 nm |
| Graphene 2 | mech. exfol. | split gates | Si/SiO$_2$ (285 nm) | dep. on $V_{g1}$, $V_{g2}$ | yes | 1.94 +/- 0.73 | 60 nm |
| Graphene 3 | mech. exfol. | single gate | Si/SiO$_2$ (285 nm) | dep. on $V_g$ | yes | 2.65 +/- 0.09 | - |
| Graphene 4 | mech. exfol. | single gate | Si/SiO$_2$ (285 nm) | dep. on $V_g$ | yes | 2.82 +/- 0.58 | - |
| Graphene 5 | mech. exfol. | single gate | Si/SiO$_2$ (285 nm) | dep. on $V_g$ | yes | 2.40 +/- 1.79 | - |
| Graphene 6 | mech. exfol. | single gate | Si/SiO$_2$ (285 nm) | dep. on $V_g$ | yes | 1.03 +/- 1.04 | - |
| Graphene 7 | CVD grown | none | SiC | 10 × 10$^{12}$ cm$^{-2}$ | no | - | - |

**Supplementary Table 1. Overview of samples as shown in Supplementary Figure 2.** Graphene 1 corresponds to the sample shown in the main manuscript.

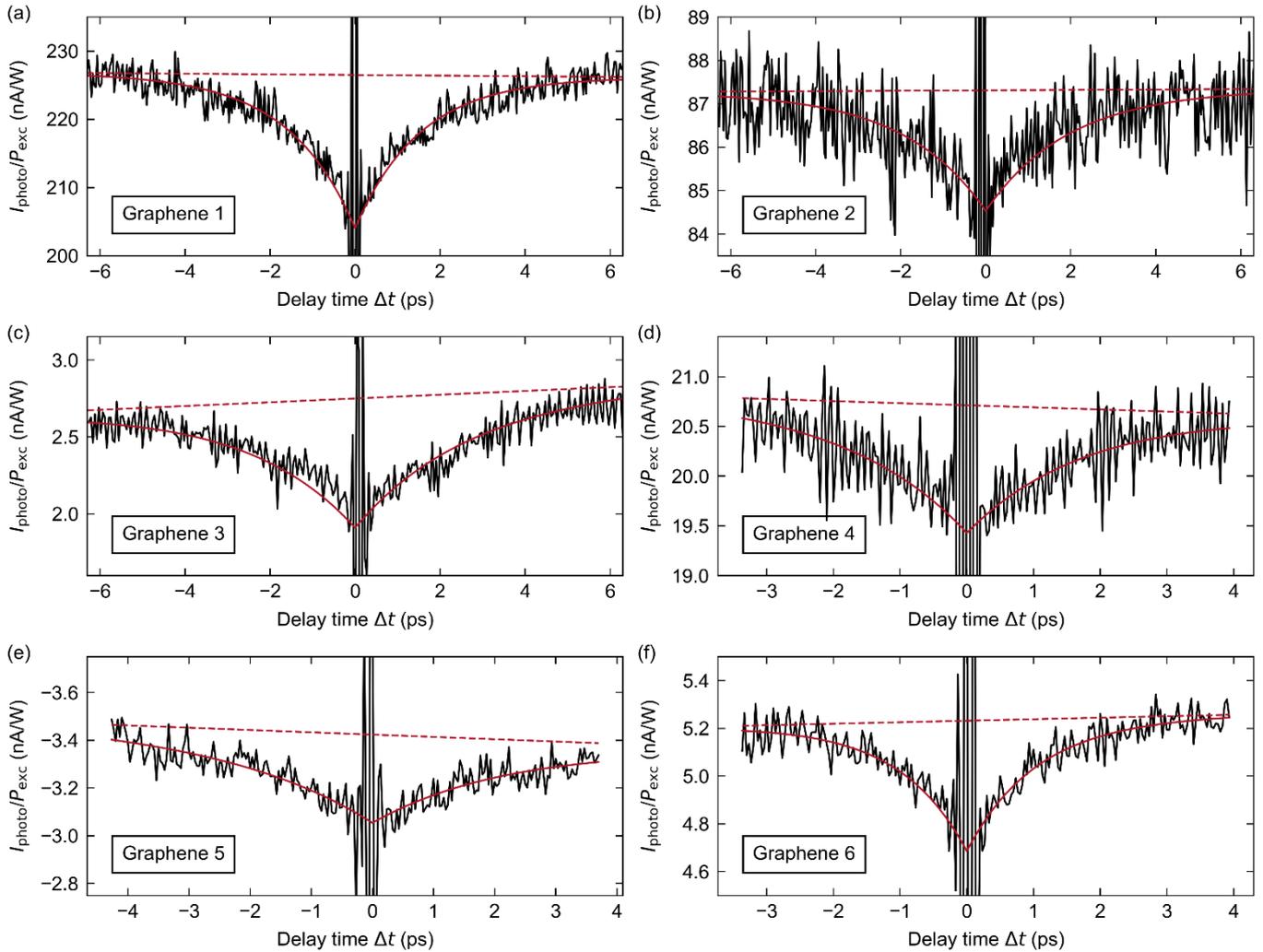

**Supplementary Figure 4. Overview of samples as defined in Supplementary Table 1.** Ultrafast photo response in (a) sample Graphene 1 (referred to as graphene sample in the main manuscript), (b) sample Graphene 2, (c) sample Graphene 3, (d) sample Graphene 4, (e) sample Graphene 5, (f) sample Graphene 6. Corresponding exponential fits (solid line) and linear backgrounds (dashed line) are shown in red.

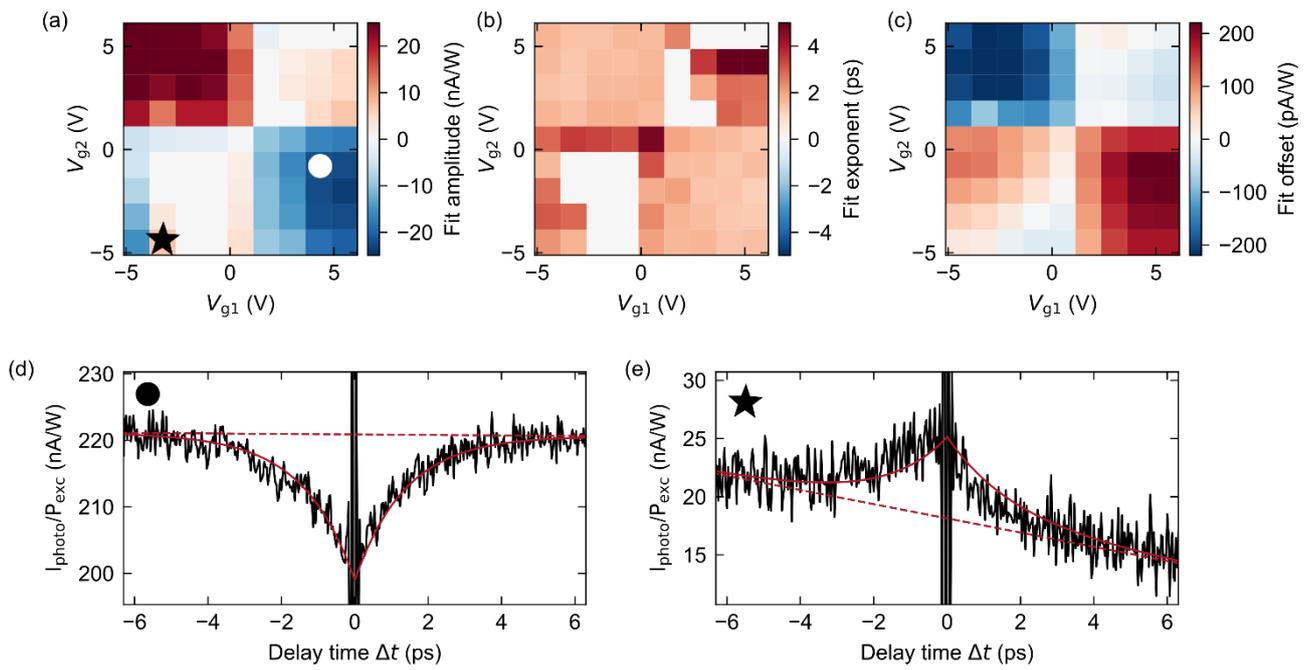

**Supplementary Figure 5. Dip- and peak-shaped photocurrent with time delay for λ = 6 μm.** (a) Amplitude, (b) time constant, (c) offset as shown in Fig. 2. The circle and the star indicate the gate voltages ($V_{g1}$ and $V_{g2}$) where the ultrafast photocurrent measurements (d) and (e) are recorded.

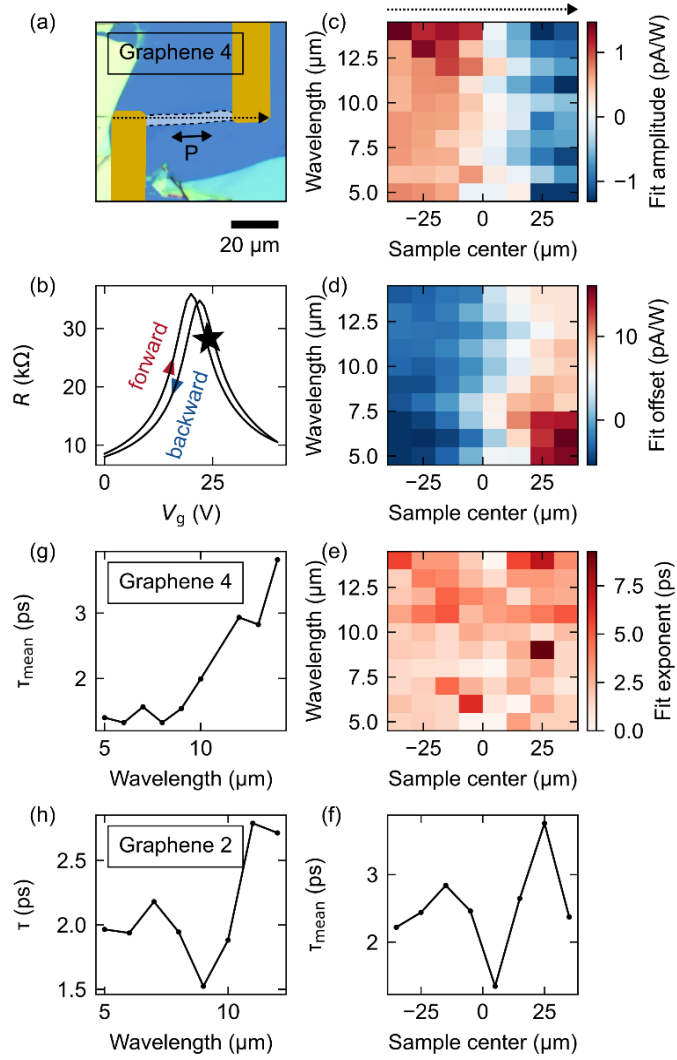

**Supplementary Figure 6. Spatial dependence for sample Graphene 4 (single gate graphene device).** (a) Microscope image of the graphene device. The laser polarization $P$ and the scan direction (dashed arrow) are indicated. (b) Transfer curve. The gate voltage used for the following measurements is indicated with a star. (c), (d), (e) Scan across the sample for different wavelengths. At each point, an ultrafast photocurrent measurement is taken and the exponential fits according to equation (1) of the main manuscript shown as maps: (c) amplitude $A(\Delta t = 0)$, (d) offset $I_{\text{offset}}$, (e) exponent $\tau$. (f) Projection of the fit exponent in (e) onto the scan position axis. (g) Projection of the fit exponent in (e) onto the wavelength axis. (h) Wavelength dependence of the time constant for sample Graphene 2 (split gate), as a comparison.

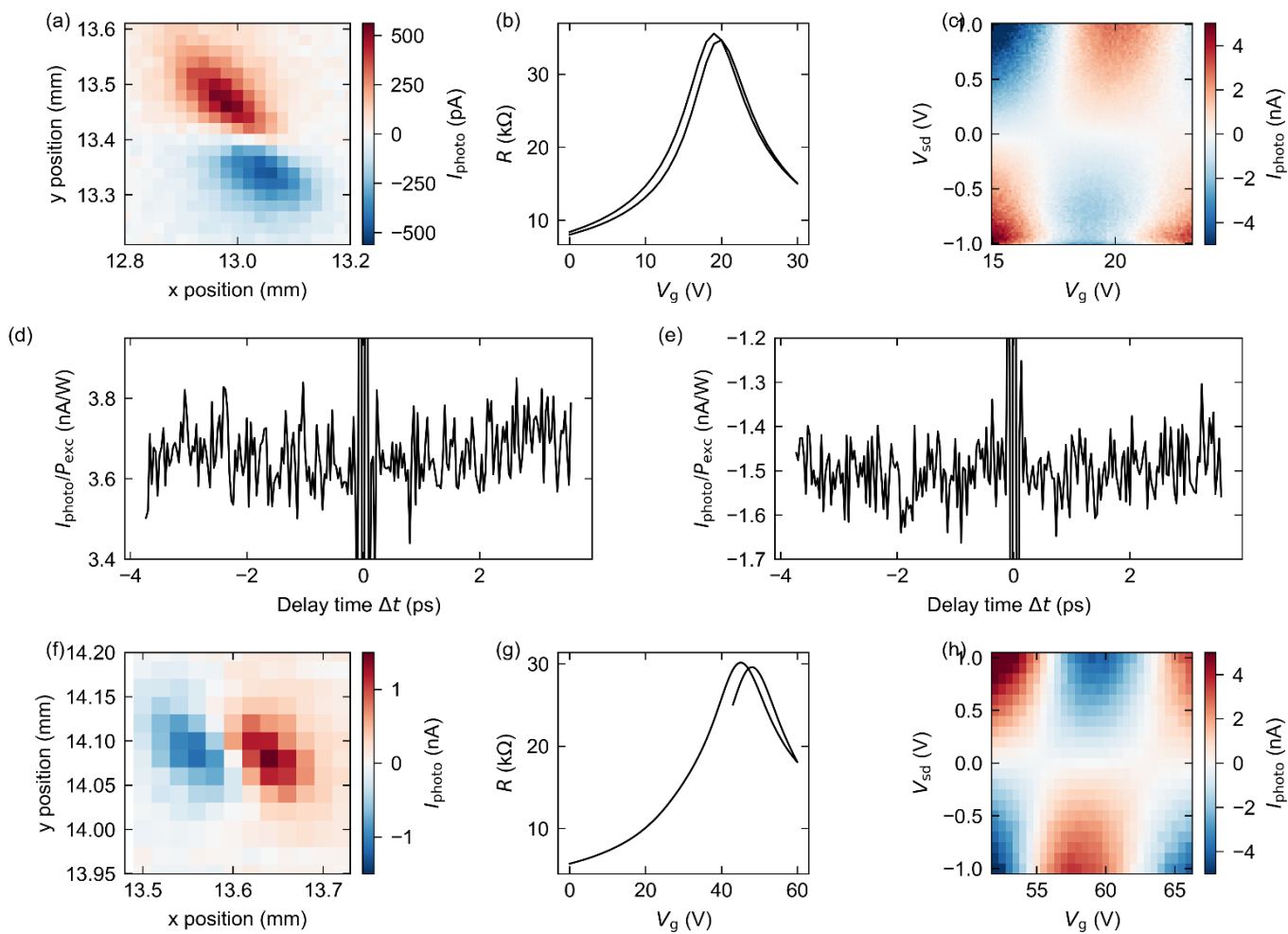

**Supplementary Figure 8. Single-gate samples in the regime of a bolometric photoresponse.** (a) Spatially resolved time-integrated photocurrent map for sample Graphene 4 at $V_{sd}$ = 0 V. (b) Transfer curve for sample Graphene 4. (c) Gate-Bias photocurrent map for sample Graphene 4. (d) Ultrafast photocurrent measurement for charge neutrality point CNP – 10 V. (e) Ultrafast photocurrent measurement for charge neutrality point CNP + 10 V. (f)-(h) Reproduced (a)-(c) for sample Graphene 6.

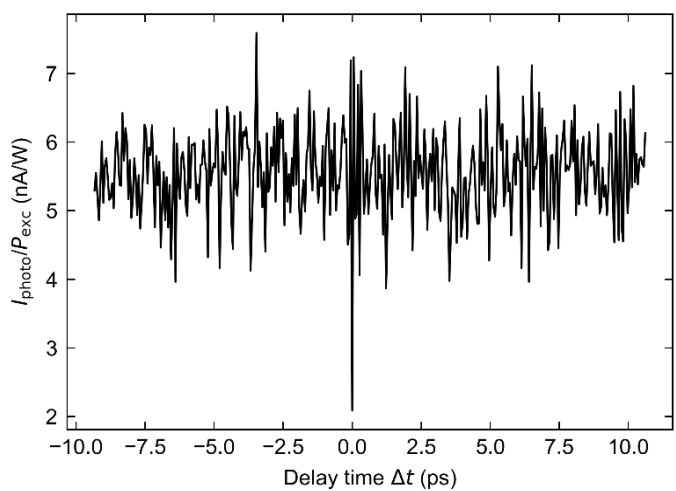

**Supplementary Figure 9.** Ultrafast photocurrent measurement for sample Graphene 7 (CVD on SiC).

**Supplementary Note 1. Transfer Matrix Method (TMM) model.**

We model the propagation of light through stratified, planar multilayer structures using the Transfer Matrix Method (TMM). At each interface between layers, continuity of the tangential components of the electric and magnetic fields is enforced via Fresnel boundary conditions. Each individual layer of refractive index $n'_j = n_j + i\kappa_j$[3–6] and thickness $d_j$ is represented by a characteristic matrix that relates field amplitudes on either side. The graphene layer is modeled by a surface conductivity. The overall response of the structure is obtained by multiplying the matrices to form the total system matrix. From this system matrix, reflection and transmission coefficients are calculated[7,8].

**Supplementary Note 2. Model Hamiltonian and charge transport approach.**

We model the graphene system by a tight-binding model including the electron-phonon coupling (EPC) based on the Holstein-Peierls Hamiltonian

$$\hat{H} = \sum_{\langle M,N \rangle} \gamma \hat{c}_M^\dagger \hat{c}_N + \sum_{\langle M,N \rangle \mathbf{Q}} \hbar\omega_\mathbf{Q} g_{MN}^\mathbf{Q} (\hat{b}_\mathbf{Q}^\dagger + \hat{b}_{-\mathbf{Q}}) \hat{c}_M^\dagger \hat{c}_N + \sum_\mathbf{Q} \hbar\omega_\mathbf{Q} \left( \hat{b}_\mathbf{Q}^\dagger \hat{b}_\mathbf{Q} + \tfrac{1}{2} \right), \quad (S1)$$

where $\gamma$ is the nearest-neighbor transfer integral of 2.7 eV, $\hbar\omega_\mathbf{Q}$ are the phonon energies with the combined index $\mathbf{Q} = (\mathbf{q}, \lambda)$ with the phonon wave vector $\mathbf{q}$ and mode index $\lambda$. The electron-phonon coupling (EPC) is described by the interatomic coupling constants $g_{MN}^\mathbf{Q}$. Using the AB-sublattice coordinates of the primitive unit cell of the honeycomb lattice, the Hamiltonian can be rewritten as

$$\hat{H} = \sum_{\langle i,j \rangle} \gamma (\hat{c}_{iA}^\dagger \hat{c}_{jB} + H.c.) + \sum_{\langle i,j \rangle \mathbf{Q}} \hbar\omega_\mathbf{Q} (\hat{b}_\mathbf{Q}^\dagger + \hat{b}_{-\mathbf{Q}})(g_{iAjB}^\mathbf{Q} \hat{c}_{iA}^\dagger \hat{c}_{jB} + H.c.) + \sum_\mathbf{Q} \hbar\omega_\mathbf{Q} \left( \hat{b}_\mathbf{Q}^\dagger \hat{b}_\mathbf{Q} + \tfrac{1}{2} \right), \quad (S2)$$

which is most convenient for the discussion of the model for the EPC below.

Instead of using the full microscopic expressions with a large number of parameters, we introduce a compact model form for the wave-vector dependence of the coupling constants $g_{iAjB}^\mathbf{Q}$:

$$g_{iAjB}^\mathbf{Q} = \frac{g_{iAjB}^{\mathbf{q}=0,\lambda}}{2\sqrt{N_\Omega}} \left( \alpha_\lambda e^{i\mathbf{q}\cdot\mathbf{R}_i} + \beta_\lambda e^{i\mathbf{q}\cdot\mathbf{R}_j} \right), \quad (S3)$$

The coefficients $g_{iAjB}^{\mathbf{q}=0,\lambda}$, $\alpha_\lambda$, and $\beta_\lambda$ describe the particular form of the mode-dependent coupling, which will be related to previous models of the EPC in graphene in literature below.[9,10] We further exploit their translational invariance, i.e., using $g_{iAjB}^\mathbf{Q} = g_{0A(j-i)B}^\mathbf{Q} = g_{0AlB}^\mathbf{Q}$ and express the Hamiltonian in the electronic momentum coordinates, using the transformation

$$\hat{c}_{iA} = \frac{1}{\sqrt{N_\Omega}} \sum_\mathbf{k} e^{-i\mathbf{k}\cdot\mathbf{R}_i} \hat{c}_{\mathbf{k}A}, \quad (S4)$$

$$\hat{c}_{jB} = \frac{1}{\sqrt{N_\Omega}} \sum_\mathbf{k} e^{-i\mathbf{k}\cdot\mathbf{R}_j} \hat{c}_{\mathbf{k}B}, \quad (S5)$$

which yields

$$\hat{H} = \sum_{\mathbf{k}l} \gamma (\hat{c}_{\mathbf{k}A}^\dagger \hat{c}_{\mathbf{k}B} e^{i\mathbf{k}\cdot\mathbf{R}_{0l}} + H.c.) + \sum_{\mathbf{Q}\mathbf{k}l} \hbar\omega_\mathbf{Q} (\hat{b}_\mathbf{Q}^\dagger + \hat{b}_{-\mathbf{Q}}) \left( \hat{c}_{\mathbf{k}A}^\dagger \hat{c}_{\mathbf{k}+\mathbf{q}B} \frac{g_{iAjB}^{\mathbf{q}=0,\lambda}}{2\sqrt{N_\Omega}} \left( \alpha_\lambda e^{i(\mathbf{k}+\mathbf{q})\cdot\mathbf{R}_{0l}} + \beta_\lambda e^{i\mathbf{k}\cdot\mathbf{R}_{0l}} \right) + H.c. \right) +$$
$$\sum_\mathbf{Q} \hbar\omega_\mathbf{Q} \left( \hat{b}_\mathbf{Q}^\dagger \hat{b}_\mathbf{Q} + \tfrac{1}{2} \right). \quad (S6)$$

The phonon modes of interest here are the two optical Γ-point modes of graphene with mode energies at around 180 meV to 200 meV depending on the literature. These modes are referred to as longitudinal optical (LO) and transverse optical (TO) modes, and possess mode patterns either aligned with the zig-zag direction (LO) or the armchair direction (TO) at the Γ-point of the phonon Brillouin zone.[11,12] (see Supplementary Figure 10)

We evaluate the Hamiltonian at $\mathbf{q} = 0$, which is used in the calculations for the relaxation time in the following form:

$$\hat{H} = \sum_\mathbf{k} \gamma (\hat{c}_{\mathbf{k}A}^\dagger \hat{c}_{\mathbf{k}B} (1 + e^{i\mathbf{k}\cdot\mathbf{a}_1} + e^{i\mathbf{k}\cdot\mathbf{a}_2}) + H.c.) + \sum_\mathbf{k} \hbar\omega_{\text{LO}} g_{\text{LO}} (\hat{b}_{\mathbf{q}=0}^\dagger + \hat{b}_{-\mathbf{q}=0})(\hat{c}_{\mathbf{k}A}^\dagger \hat{c}_{\mathbf{k}B} (e^{i\mathbf{k}\cdot\mathbf{a}_1} - e^{i\mathbf{k}\cdot\mathbf{a}_2}) + H.c.) +$$
$$\hbar\omega_{\text{LO}} \left( \hat{b}_{\mathbf{q}=0}^\dagger \hat{b}_{\mathbf{q}=0} + \tfrac{1}{2} \right), \quad (S7)$$

for the LO mode and

$$\hat{H} = \sum_{\mathbf{k}} \gamma \left( \hat{c}_{\mathbf{k}A}^\dagger \hat{c}_{\mathbf{k}B} \left(1 + e^{i\mathbf{k}\cdot\mathbf{a}_1} + e^{i\mathbf{k}\cdot\mathbf{a}_2}\right) + H.c. \right) + \sum_{\mathbf{k}} \hbar\omega_{\mathrm{TO}} g_{\mathrm{TO}} (\hat{b}_{\mathbf{q}=0}^\dagger + \hat{b}_{-\mathbf{q}=0}) \left( \hat{c}_{\mathbf{k}A}^\dagger \hat{c}_{\mathbf{k}B} \left(1 - \frac{e^{i\mathbf{k}\cdot\mathbf{a}_1}}{2} - \frac{e^{i\mathbf{k}\cdot\mathbf{a}_2}}{2}\right) + H.c. \right) + \hbar\omega_{\mathrm{TO}} \left( \hat{b}_{\mathbf{q}=0}^\dagger \hat{b}_{\mathbf{q}=0} + \frac{1}{2} \right) \quad (S8),$$

for the TO mode, where $\mathbf{a}_1$ and $\mathbf{a}_2$ are the lattice vectors. The effective modes are chosen to have mode energies $\hbar\omega_{\mathrm{LO}} = \hbar\omega_{\mathrm{TO}} = 189$ meV and moderate dimensionless coupling constant of $g_{\mathrm{LO}} = g_{\mathrm{TO}} = 0.159$ yielding an interaction strength of $\hbar\omega_{\mathrm{LO}} g_{\mathrm{LO}} = \hbar\omega_{\mathrm{TO}} g_{\mathrm{TO}} = 30$ meV. These couplings at the Dirac point result in reorganization energies of $\Lambda_{\mathrm{LO}} = 3\hbar\omega_{\mathrm{LO}} g_{\mathrm{LO}}^2 = 14.3$ meV and $\Lambda_{\mathrm{TO}} = \frac{9}{4}\hbar\omega_{\mathrm{TO}} g_{\mathrm{TO}}^2 = 10.7$ meV, respectively.

The Hamiltonians in Eqs. (S7) and (S8) have been implemented in a highly efficient charge-transport code[13–15], which we use to calculate the carrier relaxation time, that in turn is based on the dc-conductivity

$$\sigma_{\alpha\alpha}^{\mathrm{dc}}(T) = \frac{e^2}{V}\frac{\beta}{2} \lim_{t\to\infty} \frac{d}{dt} \Delta x_\alpha^2(t), \quad (S9)$$

with the thermally averaged mean-square displacement (MSD)

$$\Delta x_\alpha^2(t) = \mathrm{Tr}(\hat{\rho}\Delta \hat{x}_\alpha^2(t)), \quad (S10)$$

where $\hat{\rho}$ is the grand-canonical density operator $\hat{\rho} = \frac{e^{-\beta(\hat{H}-\mu\hat{N})}}{Z_{\mathrm{gc}}}$.

At zero temperature and in the non-interacting case, the dc-conductivity is then given by[16]

$$\sigma_{\alpha\alpha}^{\mathrm{dc},0} = \lim_{T\to 0} \sigma_{\alpha\alpha}^{\mathrm{dc},0}(T) = \frac{e^2 D(E) v_{\mathrm{F}}^2}{2} \tau_0(E), \quad (S11)$$

where $D(E)$ is the density of states of graphene and $v_{\mathrm{F}}$ is the Fermi velocity $v_{\mathrm{F}} = \frac{3}{2}\gamma a_0$, with the C-C distance of $a_0 = 1.42$ Å. To obtain the relaxation time of the interacting system, we thus may define it similarly to the non-interacting case via:

$$\lim_{T\to 0} \sigma_{\alpha\alpha}^{\mathrm{dc}}(T) = \frac{e^2 D(E_{\mathrm{F}}) v_{\mathrm{F}}^2}{2} \tau(E), \quad (S12)$$

which allows to determine the relaxation time from the evaluation of $\sigma_{\alpha\alpha}^{\mathrm{dc}}(T)$. The dc-conductivity is calculated numerically for both the interacting and the non-interacting case. Dividing the dc-conductivities by the corresponding DOS $D(E)$, we obtain the relaxation time as:

$$\tau(E) = 2 \frac{\lim_{T\to 0} \sigma_{\alpha\alpha}^{\mathrm{dc}}(T)}{e^2 D(E) v_{\mathrm{F}}^2} = \frac{\lim_{t\to\infty}\frac{d}{dt}\Delta x^2(E,t)}{V D(E) v_{\mathrm{F}}^2}, \quad (S13)$$

which can be compared for both cases. In Supplementary Note 3, we discuss the results of our charge transport calculations based on the present model Hamiltonian, including the EPC to the optical modes of graphene.

**Supplementary Note 3. Polaron dressing of ballistic electrons at the time scale of super collisions**

After introducing the model Hamiltonian and transport approach to determining carrier relaxation times, we compare our theoretical results to the experimental data. In the charge-transport simulations, we calculate the expectation value of the MSD of the hot electrons (assuming no thermal phonon occupation) in the zero-phonon subspace $|0\rangle_{\mathrm{ph}}$

$$\Delta x_\alpha^2(t) = \mathrm{Tr}_{\mathrm{el}}\left( \langle 0|_{\mathrm{ph}} \hat{\rho} \Delta \hat{x}_\alpha^2(t) |0\rangle_{\mathrm{ph}} \right). \quad (S14)$$

The results are shown in Figure 4 of the main text. Fig. 4(b) displays the carrier relaxation times defined in Eq. (S13) as function of energy $E$, taken as half the experimental excitation energy (cf. Fig. 4(a)). Without EPC (black curve), the electrons propagate ballistically $\Delta x_\alpha^2(t) \propto t^2$, since no other scattering mechanisms or disorders are considered in this case, up to the total simulation time set to $\tau_{\mathrm{ref}} = 3.7$ ps here and for all subsequent simulations. This simulation time effectively acts as coherence-limiting time (decoherence time) for all subsequent simulations, where the coherent quantum dynamics would stop by inelastic scattering mechanisms that set in, such as supercollisions. For the disorder- and phonon-free limit, the MSD follows approximately $\Delta x_\alpha^2(t) \approx \frac{e^2 D(E) v_{\mathrm{F}}^2}{2}\tau_{\mathrm{ref}}^2$ yielding $\tau(E) \approx \tau_{\mathrm{ref}}$ shown as analytical reference in Fig. 4(b) (grey dashed line). The numerical results (black curve) agree with this analytic behavior, aside from small finite-size fluctuations due to the finite system size of 1.6x10⁷ electronic orbitals.

Including LO and TO optical phonons changes the picture markedly. The relaxation time (S13) (blue curve) drops sharply from 3.7 ps to about 1.6 ps near E= 0.1 eV, followed by a gradual and eventually full recovery of the reference result at around $E$ = 0.4 eV. We observe that the dynamics is ballistic at this time scale ($\Delta x_\alpha^2(t) \propto \tau_{\text{ref}}^2$), indicating that the dip is not caused by phonon-scattering but by coherent polaron dressing (Fig. 4(a)). This polaronic dressing renormalizes the band velocity (analogous to a mass enhancement) with a maximum renormalization factor of ~0.5 (given by $\tau/\tau_{\text{ref}}$). Hence the dip of the carrier relaxation time in Fig. 4(b) is thus a signature of the coherent, ballistic propagation of the dressed electron before scattering sets in. Indeed, the polaron-dressing occurs at much earlier times, specifically at the oscillation period of the optical phonons (≈20 fs). The suppression of dressing at higher energies $E$ is attributed to faster pseudospin (sublattice) precession ($\propto |\mathbf{k}|$). The velocity renormalization becomes less effective and at higher energies (~0.22 eV) and even scattering by the optical mode is so inefficient that the EPC-free ballistic limit of the MSD is fully recovered.

Improvements to the effective model of optical $\mathbf{q} = 0$ modes towards more refined sets of microscopic EPC parameters are conceivable, but the present EPC Hamiltonian model and the considered transport scenario clearly reproduce the experimental findings for the energy dependence of the carrier relaxation time. As a result, we can attribute the drop in the experimental relaxation time to energy-dependent polaron dressing due to the EPC of the optical modes.

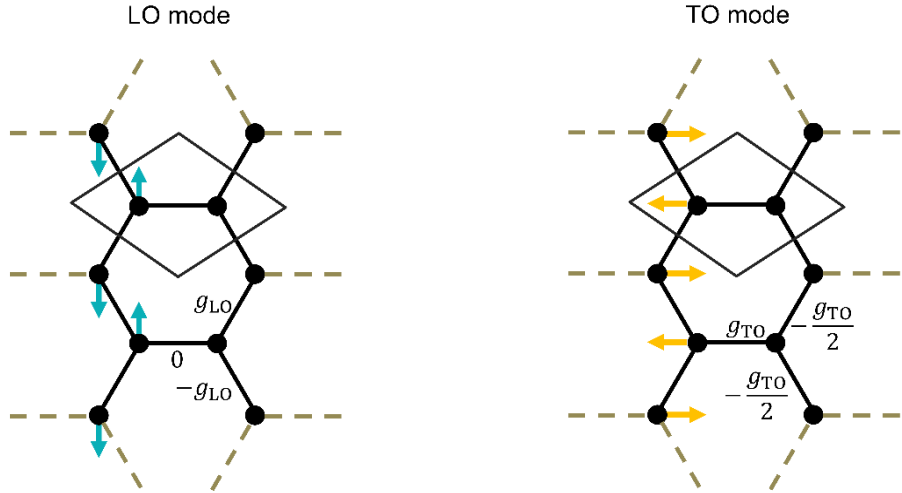

**Supplementary Figure 10.** Mode patterns of the $\Gamma$-point optical phonons for LO and TO modes in the real space. The dimensionless EPC constants $g_{LO}$ and $g_{TO}$ (see Eqs. (S7) and (S8)) follow this coupling pattern for the nearst neighbors in real space.